\documentclass[aps,prb,amsmath,amssymb,groupedaddress,10pt,superscriptaddress,floatfix,twocolumn,showpacs]{revtex4-1}
\usepackage[utf8]{inputenc}

\usepackage{graphicx}
\usepackage[colorlinks=true,linkcolor=blue,citecolor=blue,urlcolor=blue]{hyperref}
\usepackage{color}
\usepackage{amsmath}
\usepackage{siunitx}
\usepackage{xfrac}
\usepackage{braket}
\usepackage{bbm}
\usepackage{notes2bib}

\DeclareMathOperator{\tr}{Tr}
\DeclareMathOperator{\re}{Re}
\newcommand{\um}{\mathbbm{1}}

\begin{document}
\newcommand{\nvm}{$\mathrm{NV^{\mbox{-}}}\ $}
\newcommand{\nvmnosp}{$\mathrm{NV^{\mbox{-}}}$}
\newcommand{\nvz}{$\mathrm{NV^0}\ $} 
\newcommand{\nvznosp}{$\mathrm{NV^0}$}
\newcommand{\nvratio}{$\mathrm{NV^{\mbox{-}}:NV^0}\ $} 
\newcommand{\mszero}{$\mathrm{m_S = 0}\ $}
\newcommand{\msone}{$\mathrm{m_S = \pm 1}\ $}

\title{Smooth optimal quantum control for robust solid state spin magnetometry}


\author{Tobias N\"{o}bauer}
\email[]{tobias.noebauer@ati.ac.at}
\author{Andreas Angerer}
\affiliation{Vienna Center for Quantum Science and Technology and Atominstitut, TU Vienna, Stadionallee 2, 1020 Vienna, Austria}
\author{Björn Bartels}
\affiliation{Freiburg Institute for Advanced Studies, Albert-Ludwigs-Universität Freiburg, Albertstr. 19, 79104 Freiburg, Germany}
\author{Michael Trupke}
\affiliation{Vienna Center for Quantum Science and Technology and Atominstitut, TU Vienna, Stadionallee 2, 1020 Vienna, Austria}
\author{Stefan Rotter}
\affiliation{Institute for Theoretical Physics, TU Vienna, Wiedner Hauptstra\ss e 8-10, 1040 Vienna, Austria}
\author{Jörg Schmiedmayer}
\affiliation{Vienna Center for Quantum Science and Technology and Atominstitut, TU Vienna, Stadionallee 2, 1020 Vienna, Austria}
\author{Florian Mintert}
\affiliation{Freiburg Institute for Advanced Studies, Albert-Ludwigs-Universität Freiburg, Albertstr. 19, 79104 Freiburg, Germany}
\affiliation{QOLS, Blackett Laboratory, Imperial College, London, SW7 2AZ, UK}
\author{Johannes Majer}
\affiliation{Vienna Center for Quantum Science and Technology and Atominstitut, TU Vienna, Stadionallee 2, 1020 Vienna, Austria}

\date{\today}


\begin{abstract}Nitrogen-vacancy centers in diamond show great potential as magnetic\cite{balasubramanian_nanoscale_2008, maze_nanoscale_2008}, electric\cite{dolde_electric-field_2011} and thermal sensors\cite{kucsko_nanometre-scale_2013} which are naturally packaged in a bio-compatible material. In particular, NV-based magnetometers combine small sensor volumes with high sensitivities under ambient conditions. The practical operation of such sensors, however, requires advanced quantum control techniques that are robust with respect to experimental and material imperfections, control errors, and noise. Here, we present a novel approach that uses Floquet theory to efficiently generate smooth and simple quantum control pulses with tailored robustness properties. We verify their performance by applying them to a single NV center and by characterising the resulting quantum gate using quantum process tomography. We show how the sensitivity of NV-ensemble magnetometry schemes can be improved by up to two orders of magnitude by compensating for inhomogeneities in both the control field and the spin transition frequency. Our approach is ideally suited for a wide variety of quantum technologies requiring high-fidelity, robust control under tight bandwidth requirements, such as spin-ensemble based memories involving high-Q cavities\cite{kubo_hybrid_2011,amsuss_cavity_2011}.
\end{abstract}

\maketitle
\begin{figure*}[tb]
  \centering
    \includegraphics[width=1.0\textwidth]{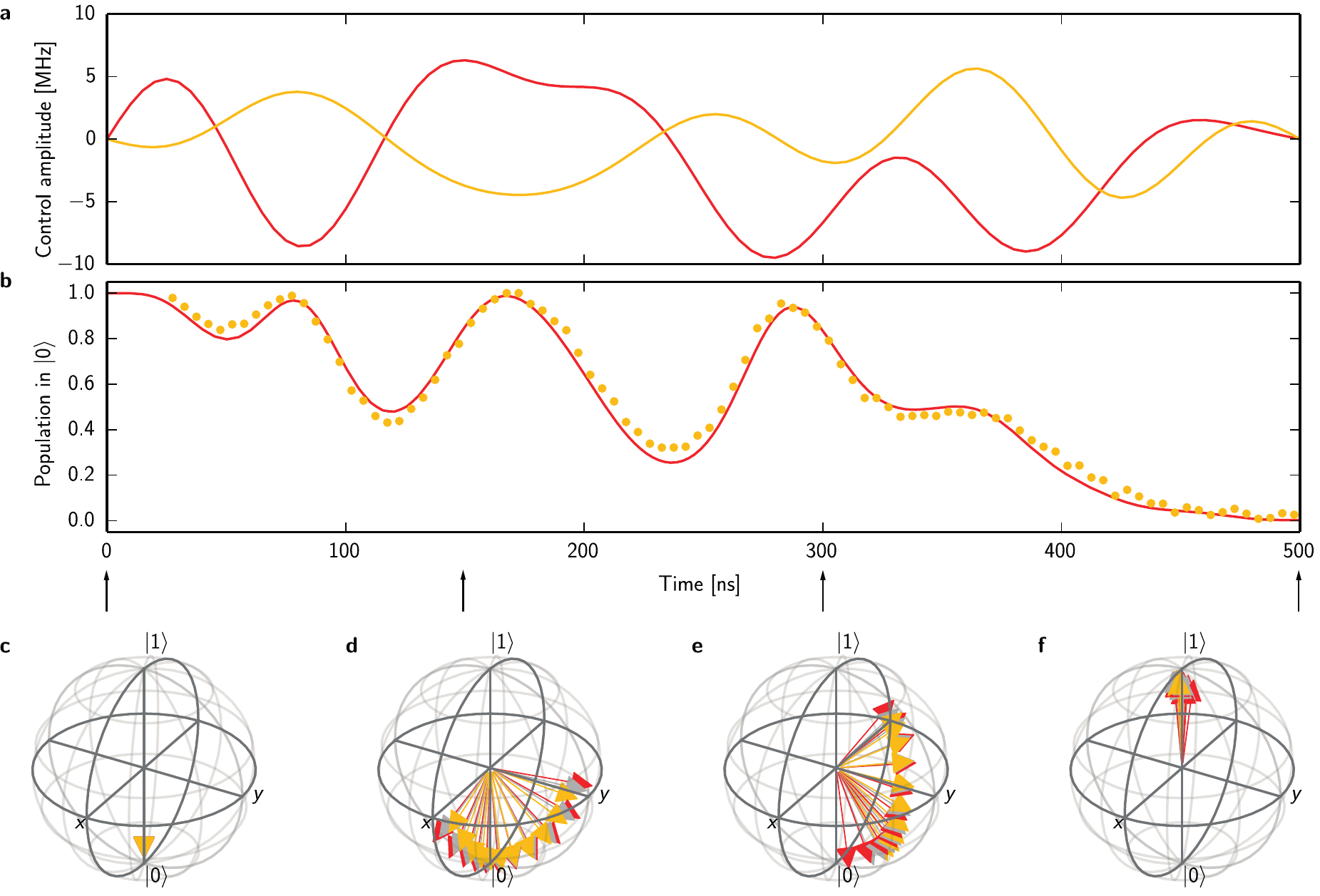}
    \caption{\label{fig1}\textbf{Qubit evolution during a smooth optimal control pulse.} \textbf{a} In-phase (red line) and quadrature (orange line) control amplitude components of a smooth control pulse implementing a robust state transfer from $\Ket{0}$ to $\Ket{1}$. The pulse consists of 10 harmonics (\num{1}, \num{2}, \ldots, \SI{10}{\MHz}), and uses a maximum control amplitude (Rabi frequency) of \SI{9.5}{\MHz}. It is optimized to be robust with respect to a $\SI{\pm 25}{\percent}$ variation in control amplitude and a Gaussian distribution of detunings of FWHM \SI{8}{\MHz}. \textbf{b} Simulated (solid line) and measured (dots) evolution of the qubit population in $\Ket{1}$ during playback of the pulse shown in (a). The simulation shows coherent evolution in the rotating wave approximation. \textbf{c-f} Snapshots of simulated Bloch vector evolutions at time offsets 0, 150, 300 and \SI{500}{\ns} during playback of the pulse (indicated by arrows) for three different control amplitude scalings (red: \SI{90}{\percent}, grey: \SI{100}{\percent}, orange: \SI{110}{\percent}) and eleven different qubit detunings (equally spaced in a $\SI{\pm 7}{\MHz}$ range; not labelled) for each of the scalings.}
\end{figure*}

NV-based sensing schemes rely on the optical detection of magnetic resonance to read out magnetic, electric or temperature-induced Larmor frequency shifts of the electron spin associated with the defect. Most commonly, rectangular microwave pulses are used to drive simple spin-echo or Ramsey-type sequences. In more recent experiments, composite pulses have been shown to improve single-NV magnetic sensitivity\cite{aiello_composite-pulse_2013} by dynamically decoupling the spin from the dephasing environment. Over the last years, great effort has been invested in optimizing measurement parameters and readout methods\cite{acosta_broadband_2010, dreau_avoiding_2011, fedder_towards_2011} and in improving materials for both single-NV\cite{balasubramanian_ultralong_2009,pham_enhanced_2012} and ensemble-based approaches\cite{acosta_diamonds_2009,nobauer_creation_2013} in bulk and nanodiamonds\cite{trusheim_scalable_2014}. 

For sensing architectures based on NV ensembles, high defect densities are in principle desirable in order to increase the fluorescence yield, but they come at the expense of increased inhomogeneous broadening due to irradiation damage in the host crystal, as well as of a decreased optical readout contrast\cite{kim_electron_2012}. In scenarios involving wide-field magnetic imaging using ensembles of NVs\cite{pham_magnetic_2011,le_sage_optical_2013} a number of additional complications arise: Inhomogeneities in the control field amplitude as well as detunings due to inherent or extrinsic inhomogeneous broadening (which can reach several \si{\MHz} in high-density samples) reduce the control fidelity. These complications become particularly severe for randomly oriented NVs in nanodiamonds. Simple rectangular pulses are therefore no longer sufficient to perform high-fidelity quantum control (such as $\pi$ or $\frac{\pi}{2}$ pulses on the NV spins) and to accurately map induced spinor phase shifts to population differences. In bulk crystals, the dominating source of control field inhomogeneities are the varying distances of the spins from the current densities (in wires or antennas) that excite the control field. For example, when applying microwaves using a straight wire of rectangular cross section ($\num{100} \times \SI{20}{\um}$) placed on top of a diamond sample containing shallow NVs, we measure a control field amplitude decrease of one third within a distance of \SI{60}{\micro\meter} from the wire (see Methods).

Here, we demonstrate an efficient and robust quantum control approach that is able to greatly improve operator fidelities (and hence sensitivities) in such challenging scenarios and with limited resources (such as bandwidth and power). The effectiveness of (quantum-) optimal control algorithms\cite{brif_control_2010} such as Krotov\cite{krotov_global_1995}, GRAPE\cite{skinner_application_2003} and CRAB\cite{doria_optimal_2011} for pulse shaping has been demonstrated in a number of fields, ranging from NMR\cite{skinner_application_2003} to ultracold atoms\cite{bucker_vibrational_2013}. Recently, similar techniques were shown to improve fidelities of strongly driven single-NV spin flips\cite{scheuer_precise_2013} and entangling gates in NV-nuclear spin quantum registers\cite{dolde_high_2013}. The advantage of our method of choice\cite{bartels_smooth_2013} is that it permits to easily construct spectrally narrow pulses within a well-defined frequency window. In practice, we parametrize the control pulse in terms of a small set (typically, 8 to 10) of time-periodic functions, and perform variational analysis in Floquet space to optimize these parameters to meet robustness and fidelity requirements under contraints. 

For magnetometry with high-density, inhomogeneously broadened NV ensembles, we optimize for robustness with respect to control amplitude variations of \SI{+- 25}{\percent} and detunings (inhomogeneous linewidth) as high as $\SI{+- 4}{\MHz}$. We constrain the control amplitudes occurring in the pulse to the maximum Rabi frequency achievable using our equipment, i.e. \SIrange{10}{20}{\MHz}. A typical resulting control pulse is shown in Fig.~\ref{fig1}(a) (solid line): Consisting of only ten frequency components with constant in-phase and quadrature amplitude coefficients, this pulse is designed to implement a state transfer from state $\Ket{0}$ to $\Ket{1}$ with a theoretical infidelity not larger than one percent within the robustness windows specified above. As illustrated in Fig.~\ref{fig1}(c), all spins in this parameter range are rephased in the target state at the end of the pulse. We note that our technique allows for incorporating a variety of other design goals, such as minimal pulse duration\cite{bartels_smooth_2013}. More details on our theoretical approach can be found in the Supplementary Information.

In order to verify its properties, we now apply the pulse shown in Fig.~\ref{fig1}(a) to an effective qubit consisting of the $\Ket{m_S=0}$ and $\Ket{m_S=-1}$ states of a single NV by modulating the two control amplitude components onto a resonant carrier microwave. Since the hyperfine structure caused by the $\mathrm{^{14}N}$ nuclear spin associated with the NV would have complicated the dynamics, we polarize the nuclear spin by optical pumping in a static magnetic field of \SI{\approx. 50}{mT} aligned with the NV symmetry axis\cite{jacques_dynamic_2009}. We interrupt the pulse at a series of time offsets and optically read out the resulting spin projection of the NV, resulting in the data shown in Fig.~\ref{fig1}(b). We find very good agreement with the expected behaviour as simulated using a simple two-level Hamiltonian\bibnote[suppl]{Further details are included in the supplementary information available online.}.
\begin{figure*}[tb]
  \centering
    \includegraphics[width=1.0\textwidth]{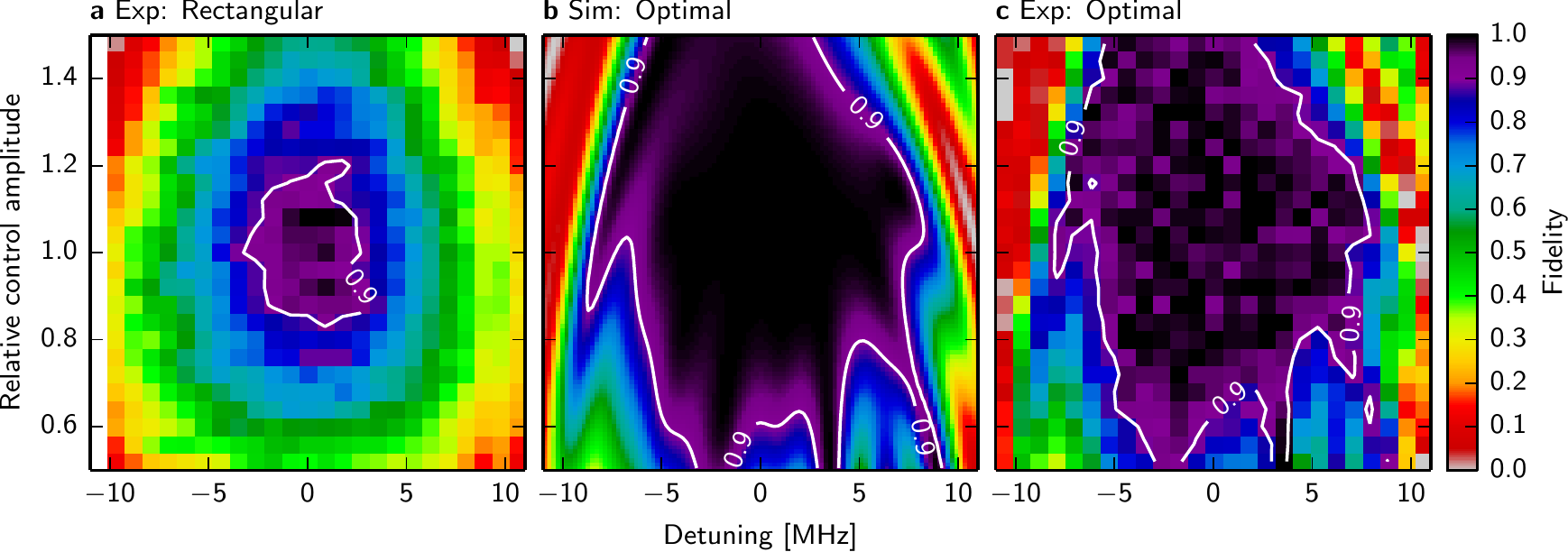}
    \caption{\label{fig2}\textbf{State transfer fidelity of rectangular and optimal pulses compared.} \textbf{a} Measured state transfer fidelity for a rectangular pulse using the same maximum control amplitude as the optimal pulse. \textbf{b} Simulated and \textbf{c} measured state transfer fidelity for a range of detunings and control amplitude scalings for the optimal pulse shown in Fig.~\ref{fig1}(a). White lines are contour lines at fidelity 0.9. The range of detunings shown here corresponds to \SI{\approx \pm 110}{\percent} of the maximum resonant Rabi frequency used in the pulses (\SI{9.5}{\MHz}). Since optical spin readout is limited by photon shot noise, exceedingly long integration times would be required to directly observe the very high theoretical fidelities in the experiment. The shape of the fidelity landscape however gets reproduced clearly in the data.}
\end{figure*}

Using the same polarization and readout scheme, we now proceed to experimentally verify that the pulse is indeed robust with respect to the detuning range (inhomogeneous broadening) and variations of the control amplitude that were required in the design procedure. To do so, we apply the full pulses and scan the carrier frequency over a range of \SI{+- 10}{\MHz} across resonance and vary the control amplitude by \SI{+- 50}{\percent} with respect to the central value for which the pulse was designed. The result is shown in Fig.~\ref{fig2} in comparison to the theoretical fidelity landscape, which is in excellent agreement with the data, only limited by the photon shot noise inherent to the optical spin readout. For comparison, we performed simulation and experiment with a rectangular pulse of the same maximum control amplitude. It is apparent that our optimal control approach is able to meet the demanding robustness requirements relevant for NV-ensemble-based sensing and results in pulses with greatly increased fidelities over a much wider range of detunings (as much as \SI{\pm 80}{\percent} of the resonant Rabi frequency) and control amplitudes (up to \SI{\pm 40}{\percent} of the correct value). Our optimal pulses also outperform basic composite pulses\cite{levitt_composite_2007} such as the $\frac{\pi}{2}_y - \pi_x - \frac{\pi}{2}_y$ and other variable rotation sequences\bibnotemark[suppl].

\begin{figure*}[tb]
  \centering
    \includegraphics[width=1.0\textwidth]{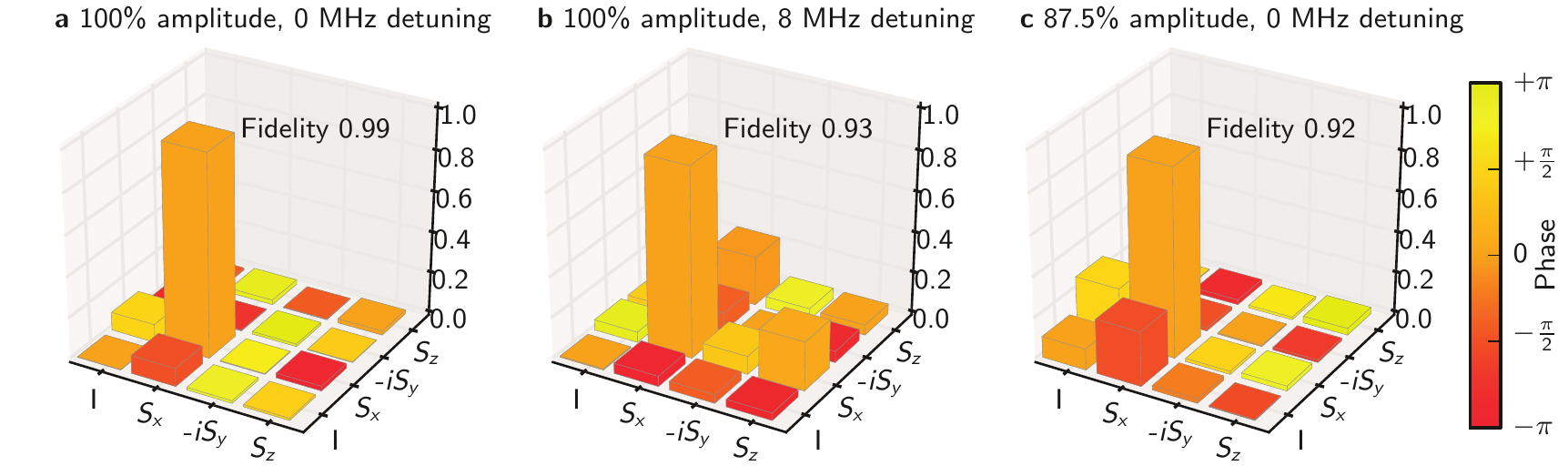}
  \caption{\label{fig3}\textbf{Robustness of single qubit gate implemented by smooth optimal control pulse characterized by quantum process tomography.} Process tomography of a $\pi_x$ smooth optimal control pulse for \textbf{a} correct ($\SI{100}{\percent}$) control amplitude and no detuning, \textbf{b} for correct control amplitude and $\SI{8}{\MHz}$ detuning, \textbf{c} for $\SI{87.5}{\percent}$ of the correct control amplitude and no detuning. Maximum resonant Rabi frequency used in the pulse: \SI{20}{\MHz}. See Supplementary Information for measurement scheme and analysis details.}
\end{figure*}

In the experiments discussed so far, we prepared our qubit in a well-defined initial state ($\Ket{m_S=0}$) before applying the control pulses, so the pulses we examined merely transferred the system from one specific initial state to one final state. However, in interferometric sensing schemes (based on Ramsey or spin echo sequences) as well as in general quantum information processing tasks, it is imperative that the pulses implement a specific time evolution, i.e. propagator, and not just a transfer from one specific state to another (which could be achieved using many different propagators). Experimentally, this property can be proven by performing quantum process tomography\cite{hohenester_optimal_2007}: Each member of a complete set of Hilbert space basis states is prepared, the operation under test applied to each, and the result characterized by state tomography\bibnotemark[suppl]. When performing this experiment with a $\pi_x$ pulse on resonance and with ideal control amplitude, we obtain a process matrix that agrees with the ideal result at a fidelity of \SI{99}{\percent} (see Fig.~\ref{fig3}). Remarkably, when repeating the tomography with the same pulse scaled to \SI{87.5}{\percent} of the ideal amplitude, the fidelity decreases only slightly to a value of \SI{92}{\percent}. When using the ideal control amplitude, but a detuning of \SI{8}{\MHz}, we measure a fidelity of \SI{93}{\percent} (Simulated values: \SI{92}{\percent} for both cases. Simulated values for rectangular pulses of same maximum Rabi frequency: \SI{85}{\percent} and \SI{84}{\percent}, respectively).

\begin{figure*}[tb]
  \centering
    \includegraphics[width=1.0\textwidth]{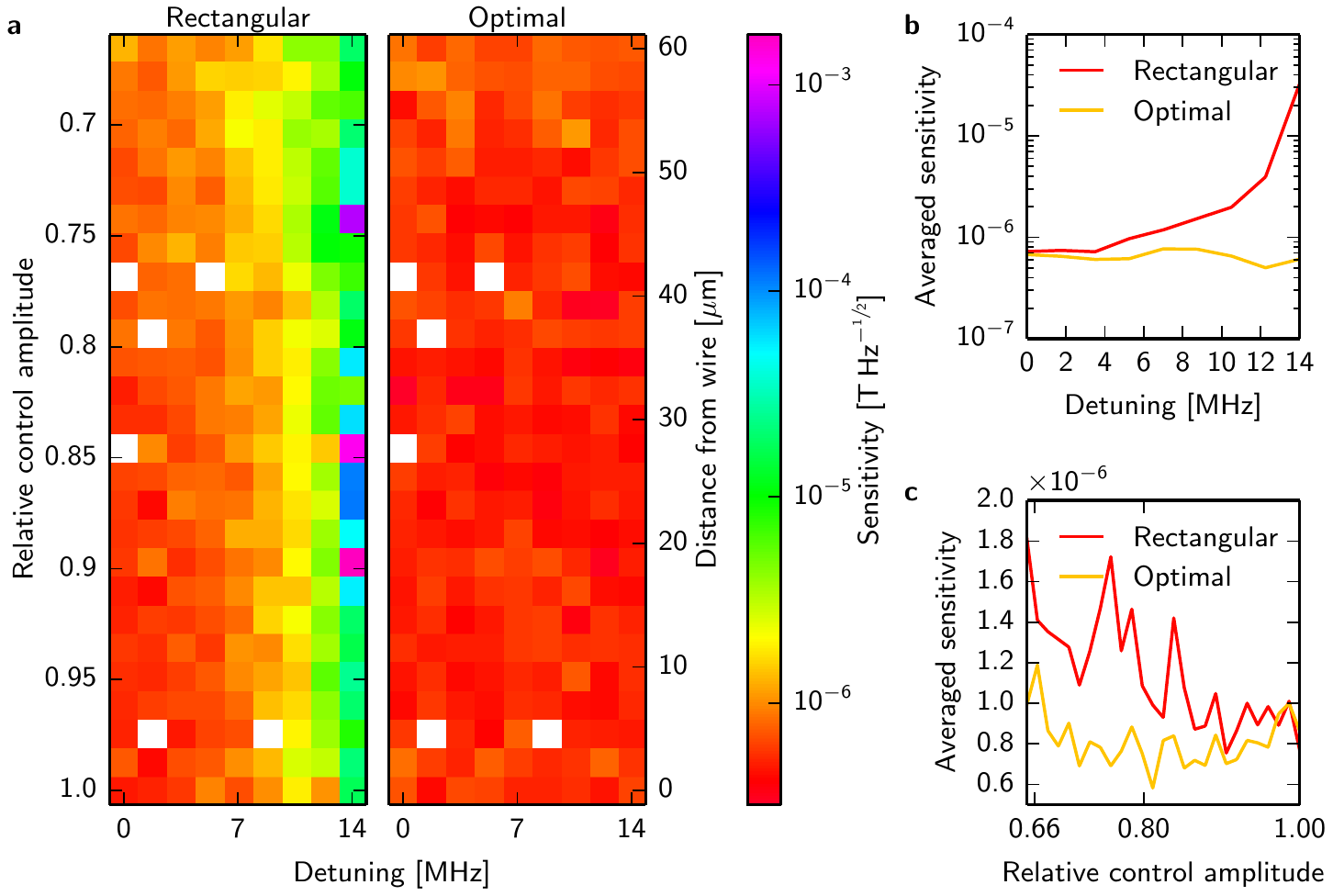}
  \caption{\label{fig4}\textbf{Improved magnetometric sensitivity under conditions akin to wide-field sensing.} \textbf{a} Measured ensemble-based sensitivity of a spin-echo AC magnetometry scheme using rectangular pulses (left panel) and smooth optimal pulses (right panel) for a range of detunings and control amplitude scalings. Note the logarithmic color scale. White pixels indicate data points that were discarded due to experimental instabilities. The smooth pulse sequence consisted of a $\Ket{0} \rightarrow \left(\Ket{0}+\Ket{1}\right)/\sqrt{2}$ state transfer pulse, followed by a phase collection time $\tau = \SI{1.2}{\us}$, a $\pi_y$  pulse, collection time $\tau$, and a final $\frac{\pi}{2}_y$ pulse. See Supplementary Information for pulse design parameters. Right vertical axis indicates approximate distances from the antenna wire ($\num{100} \times \SI{20}{\um}$) generating the control fields (data obtained from a linear fit to measurements of relative Rabi frequency vs. distance from wire) \textbf{b} Data from (a) integrated along the vertical axis and \textbf{c} horizontal axis.}
\end{figure*}

Having shown the robustness and fidelity of our smooth control pulses using a single NV center, we are now in a position to apply them to the system that they were designed for -- an ensemble of inhomogeneously broadened NVs -- and perform magnetic sensing. We use a layer of NVs (thickness \SI{8}{\nm}), which was formed \SI{12}{\nm} below the surface of a CVD diamond of high chemical purity by nitrogen ion implantation. The NV ensemble has a linewidth of \SI{960}{kHz} (FWHM). We perform AC magnetometry by repeatedly initializing the NV spins in the $\Ket{m_S=0}$ state by non-resonant optical pumping, driving a spin echo sequence with fixed free precession time $\tau$ and reading out the spin projection via optically detected magnetic resonance, akin to the protocol used in earlier work\cite{maze_nanoscale_2008}. To characterize our magnetometer, we deliberately apply an AC magnetic field of period $2\tau$ in phase with the spin echo sequence and scan its amplitude $B$. This causes a modulation of the spin echo amplitude which serves as the magnetometric signal $S(B) \propto \cos g \mu_B B$, where $g\approx2$ is the NV electron g-factor, and $\mu_B$ is the Bohr magneton. From the slope $\delta S/\delta B$ of the signal and the noise level we extract the sensitivity, which we plot in Fig.~\ref{fig4}. For comparison, we performed the experiment with simple, rectangular control pulses, as well as with optimal pulses. In order to emulate conditions typical for wide-field sensing geometries, we scan the detuning of our ensemble as well as the control amplitude across a range of values. While for the optimal pulses the sensitivity remains constant across the recorded parameter range, for the rectangular pulses it drops off by one to two orders of magnitude as the detuning becomes larger (i.e. for NVs further away from a bias wire in wide-field magnetometry, or for broad ensembles). Similarly, but less pronounced, we observe a larger drop in sensitivity at large control amplitude mismatches for the rectangular pulses than for the optimal pulses. All sensitivity values are corrected for count rate limitations of our setup\bibnotemark[suppl].

The absolute magnetic sensitivity achievable with an NV ensemble is inversely proportional to the square root\cite{taylor_high-sensitivity_2008} of its dephasing time $T_2$ and depends on a number of material parameters\cite{kim_electron_2012}. In our experiments, we have  $T_2=\SI{2.2}{\us}$, which can be greatly extended using advanced growth techniques\cite{pham_magnetic_2011}.

To summarize, we have demonstrated how the magnetic sensitivity of NV-ensemble magnetometry can be improved by up to two orders of magnitude. Our strategy builds on greatly reducing control errors due to inhomogeneous broadening and control amplitude variations using smooth, robust optimal control pulses. The absolute sensitivity achieved in our experiments is limited by the density of NVs, their dephasing time and the readout contrast, all of which can be further improved by optimizing implantation and material parameters. Using quantum process tomography, we have shown that our smooth control pulses are able to perform quantum gates at high fidelity even under significant deviations (\SI{25}{\percent}) in control amplitude and large detunings (\SI{40}{\percent} of Rabi frequency), while meeting tight bandwidth requirements. Thus, the versatility and robustness of our smooth quantum optimal control technique enable a broad range of applications across material systems.

\section*{Methods}
\subsection*{NV center as a qubit and magnetometer}
The negatively charged nitrogen vacancy center in diamond (NV) is a point defect consisting of a nitrogen atom on a carbon lattice site and an adjacent vacancy. Its optical ground state is a spin triplet, with the \mszero and \msone states split by $D=\SI{2.87}{\GHz}$ in zero external magnetic field. The spin projection can be initialized and read out optically due to a spin-selective shelving mechanism: The \msone states fluoresce up to \SI{30}{\percent} less due to an increased branching ratio for an intersystem crossing process from the optical excited state (spin triplet) to a metastable intermediate singlet state via a phonon-mediated process. From the singlet state, the system preferrably decays into the \mszero sublevel of the ground state. Hence, simple off-resonant optical excitation results in a spin-projection dependent fluorescence rate (into a broad window from \SIrange{\approx 637}{800}{\nm}) as well as spin polarization. 

In the most likely case of a center involving a $\mathrm{^{14}N}$ nucleus (nuclear spin $I=1$), the electron spin resonance transitions are hyperfine-split by $\SI{\pm 2.3}{\MHz}$. In our single-NV experiments (Figs.~1--3), we prepare our qubit level $\Ket{0} = \Ket{m_S=0, m_I=-1}$ by off-resonant optical pumping in a magnetic field of \SI{\approx 50}{mT} aligned with the NV symmetry axis. In addition to electron spin polarization, this leads to nuclear spin polarization due to spin flips in the vicinity of an excited state level anticrossing\cite{jacques_dynamic_2009}. The level $\Ket{1} = \Ket{m_S=-1, m_I=-1}$ is used as the qubit excited state. For the magnetometry experiments, the nuclear spins were not polarized, and the static magnetic bias field was \SI{\approx 3}{mT}. Any additional external magnetic fields change the Zeeman shift by $\Delta \omega \approx 2 \pi \times \SI{28}{\MHz \per \milli \tesla}$, which can be read out using the spin-echo based pulse sequence described in the main text.

\subsection*{Experimental setup and sample description}
All experiments were carried out in a home-built room temperature confocal microscope: Light from an intensity-stabilized \SI{532}{\nm} diode pumped solid state laser is chopped using a fast acousto-optical modulator with \si{\ns} rise time, passed through a single-mode optical fibre, reflected off a dichroic mirror and focussed onto the piezo-stage mounted sample using an oil immersion objective (Olympus PLAPON NA=1.4). Fluorescence from the sample is collected by the same objective, passed through the dichroic  mirror (passband \SI{> 650}{\nm}), focussed onto a confocal pinhole to block out-of-focus light and imaged onto two single photon counting modules (Perkin Elmer) through a non-polarizing beam splitter. An additional filter (passband \SI{> 750}{\nm}) is placed into the beam path after the confocal pinhole during the magnetometry experiments in order to increase the readout contrast\bibnotemark[suppl].

Microwaves from two independent sources (Anritsu and TTI) are used to drive electron spin resonance transitions. The smooth control pulses are downloaded into a home-built FPGA-based arbitrary waveform generator (200~MS/s, 14~bit, 2~ch) and amplitude-modulated onto the carrier microwaves using an IQ mixer (Marki IQ LMP-1545). A pulse generator card (Spincore Pulseblaster) is used to trigger and time the pulse sequences. Microwaves are amplified to up to \SI{30}{dBm} (Minicircuits ZHL-16W-43+) and delivered to the sample via a coplanar waveguide sample holder and a \SI{100}{\um} gold wire placed across the sample. To calibrate the decrease in control amplitude with distance from the gold wire (right vertical scale in Fig.~\ref{fig4}(a)), we perform Rabi frequency measurements at a series of distances from the wire while keeping the exciting microwave power constant. The resulting dependence is captured well by a linear fit.

Static magnetic fields of several \si{\milli \tesla} are applied using a set of three coils mounted at right angles symmetrically around the focal spot and aligned to the defect symmetry axis (crystallographic $\langle 111 \rangle$ direction). Larger static fields are applied by mounting a permanent magnet next to the sample. Small AC magnetic fields for magnetometer calibration are applied using a small coil placed next to the sample, which is driven by a commercial arbitrary waveform generator (Agilent).
For the single-NV experiments, naturally occurring defects in isotopically pure CVD diamonds (Element-6 quantum grade) were used. For ensemble-based work, a CVD diamond of natural isotope composition (Element-6 electronic grade) was implanted with nitrogen ions and subsequently annealed to form a \SI{\pm 4}{\nm} layer \SI{12}{\nm} below the surface of the sample.

\subsection*{Acknowledgements}
We gratefully acknowledge Matthew Markham of Element Six for supplying the isotopically purified diamond sample. We thank Friedrich Aumayr and his team for implanting one of the samples used with nitrogen ions, and Kathrin Buczak for characterising that sample. T. N. acknowledges the Vienna Graduate School for Complex Quantum Systems (CoQuS), funded by the Austrian Science Fund FWF which also provided financial support to J.S. (Wittgenstein Award). S.R. acknowledges financial support by the Austrian Science Fund FWF through project No. F49-P10 (SFB NextLite). F.M. and B.B. gratefully acknowledge financial support by the European Research Council within the project ODYCQUENT. A.A. acknowledges support by the Austrian Science Fund (FWF) in the framework of the Doctoral School ``Building Solids for Function'' (Project W1243).

\appendix
\section{Smooth optimal control theory}
As recently introduced\cite{bartels_smooth_2013,PhysRevA.76.052304}, smooth optimal control searches for control amplitudes $a_{jk}$ such that a certain target functional $\mathcal{F}$ becomes maximal. Here, the amplitudes $a_{jk}$ are the Fourier components in the control Hamiltonian
\begin{equation}\label{four}
H_c(t)=\sum_k\sum_{j=1}^N a_{jk}\sin(j\Omega t)\mathbf{h}_k
\end{equation}
with fundamental frequency $\Omega$ and operators $\mathbf{h}_k$. Control is exerted on a time-independent system Hamiltonian $H_0$. A common way to find the right amplitudes is to use gradient-based methods. For this purpose, one has to calculate the derivatives $\frac{\partial U}{\partial a_{jk}}$ of the time evolution operator $U$. As the Hamiltonian of the problem is periodic in time, Floquet's theorem can be used and $U$ can be computed in frequency space\cite{bartels_smooth_2013,goelman_design_1989}. The derivatives can be evaluated using perturbation theory in Floquet space, as a derivative is nothing else but the response to an infinitely small perturbation\cite{bartels_smooth_2013}.

In the present work, the case of an NV center controlled by a microwave pulse can be modelled by the Hamiltonian
\begin{equation}H=\frac{\omega_0}{2}\sigma_z+2\alpha(\mathbf{r})\cos(\omega_0 t)(f_1(t)\sigma_x+f_2(t)\sigma_y).
\end{equation}
Using the rotating-wave approximation and comparing with Eq.~\eqref{four} yields:
\begin{equation}
H_0=\frac{\omega_0}{2}\sigma_z\qquad\mathbf{h}_1=\alpha\sigma_x\qquad\mathbf{h}_2=\alpha\sigma_y.
\end{equation}
As we target the implementation of pulses that work for ensembles of NV centers with different detunings $\omega_0$ and relative control amplitudes $\alpha$, the target functionals will always be averaged over these quantities. More explicitly, we used the following target functionals:
\begin{equation}\label{fid1}
\mathcal{F}_\text{fid}=\lvert\langle\psi_f\vert U(t_f)\vert\psi_i\rangle\rvert^2,
\end{equation}
where an initial state $\vert\psi_i\rangle$ is to be transferred to a final state $\vert\psi_f\rangle$ at a moment $t_f$ in time, and the operator measures
\begin{equation}\label{fid2}
\mathcal{F}_\text{op}=\frac{1}{2}\re\tr\left(U(t_f)U_f^\dagger\right),
\end{equation}
where a unitary gate $U_f$ is to be implemented at $t=t_f$. In addition, a penalty functional
\begin{equation}\label{pen}
\mathcal{F}_p=-p\sum_{j,k}a_{jk}^2
\end{equation}
with a parameter $p>0$ can be used in order to limit the power of the pulse. In practice, we started our algorithm with a high value of $p$ and then reduced it in each iteration step by a small amount $\Delta p$. If, however, the maximal amplitude of the pulse exceeded a certain threshold $A_\text{max}$, we again increased $p$ by $\Delta p$. This method also helps to avoid local maxima, as compared to the unbounded control.

In the following, the reader will find the Fourier components (in MHz) of the pulses we used:
\begin{widetext}
\begin{center}
\begin{tabular}{ccccccccccc}
 pulse & $a_{1,1}$ & $a_{1,2}$ & $a_{1,3}$ & $a_{1,4}$ & $a_{1,5}$ & $a_{1,6}$ & $a_{1,7}$ & $a_{1,8}$ & $a_{1,9}$ & $a_{1,10}$\\
\hline 
 $\pi$ & -1.177 &   1.646  & -0.549  & -1.668  & -0.627 &   0.151  &  1.680  & -0.024 & 0.858  &  1.311\\
$(\pi/2)_y$ & 1.248  & -0.573 &  -4.553 &  -0.530 &  -8.790  &  0.677 &  -1.413 &   0.736 & -4.158  &  2.075\\
$\pi_x$ & 6.498 &  -0.916 &  -5.901  &  0.169 &   2.727 &   0.929 &  -6.137 &  -0.009 & -10.532 &  -3.960\\
\end{tabular}
\end{center}

\begin{center}
\begin{tabular}{ccccccccccc}
 pulse & $a_{2,1}$ & $a_{2,2}$ & $a_{2,3}$ & $a_{2,4}$ & $a_{2,5}$ & $a_{2,6}$ & $a_{2,7}$ & $a_{2,8}$ & $a_{2,9}$ & $a_{2,10}$\\
\hline 
 $\pi$ & -0.150  & -0.355  &   0.253 &   1.165  &  0.069 &   0.470 & -0.649  & -0.814  &  0.643  & -0.657 \\
$(\pi/2)_y$ & 6.454  & -0.904 &  -6.097 &  -0.376 &   4.378 &  -2.946 & -7.539  & -1.205 & -10.375 &   1.931\\
$\pi_x$ & 0.572  &  0.483 &  -3.400 &  -0.147  & -9.616 &  -0.126 & -0.361 &  -1.531 &  -0.649 &   1.035\\
\end{tabular}
\end{center}
\end{widetext}

as well as the other parameters of the pulse (pulse duration $T$, bandwidth $\nu_\text{max}$ and maximal Rabi frequency $A_\text{max}$):

\begin{center}
\begin{tabular}{cccc}
pulse & $T$/ns & $\nu_\text{max}$/MHz & $A_\text{max}$/MHz\\
\hline
$\pi$ & 500 & 10 & 9.49\\
$(\pi/2)_y$ & 250 & 20 & 18.8\\
$\pi_x$ & 250 & 20 & 19.4\\
\end{tabular}
\end{center}

\section{Quantum process tomography and process fidelity}
\subsection{Quantum process tomography}
In order to analyse the performance of the smooth control pulses as quantum maps (as opposed to just giving their effect on a single input state), we perform quantum process tomography (QPT) and calculate the experimental process fidelity. We implement the QPT scheme given in Ref.~\onlinecite{howard_quantum_2006}, which is summarized and applied to our case in the following:

We aim at characterising the effect of a process $\mathcal{E}$ on an arbitrary input quantum state $\rho$, $\mathcal{E}\left(\rho\right)$. We write $\mathcal{E}\left(\rho\right)$ in the $\chi$-matrix representation:
\begin{align}\label{eq:chimatrix}
\mathcal{E}\left(\rho\right)=\sum_{m,n=1}^{d^2}\chi_{mn}A_m\rho A_n^\dagger,
\end{align}
where the operators $A_i$ form a complete basis set of operators, and $d$ is the dimension of our state space. The matrix $\chi$ is positive Hermitian by construction and fully characterizes the process $\mathcal{E}$. The trace preservation constraint on $\mathcal{E}$ results in the completeness relation $\sum_{m,n=1}^{d^2}\chi_{m,n}A_n^\dagger A_m =\mathbbm1$. 

For a two-level system ($d=2$), the matrix $\chi$ has $4\times4$ elements of which four follow from the completeness relation, and twelve have to be measured. By choosing $A_i = \left\lbrace I, \sigma_x, (-i\sigma_y), \sigma_z \right\rbrace$ (where $I$ is the identity and $\sigma$ are the Pauli spin operators), the task reduces to a rather simple procedure, as given in Ref.~\onlinecite{nielsen_chuang}:
\begin{itemize}
\item The qubit is consecutively prepared in each of the four initial states $\ket{\Psi_i} = \{ \ket0$, $\ket1$, $\ket{+} = \frac{1}{\sqrt{2}}(\ket0+\ket1)$, $\ket{-} = \frac{1}{\sqrt{2}}(\ket0+i\ket1)\;\}$. NV centers can be easily prepared in $\ket0$ by off-resonant optical excitation, the other states are reached from $\ket0$ by applying (rectangular) $\pi$ and $\frac{\pi}{2}$ pulses.
\item For each initial state, the process $\mathcal{E}$ (i.e., the smooth optimal control pulse) is applied.
\item To determine the resulting states $\mathcal{E}\left(\ket{\Psi_i}\bra{\Psi_i}\right)$ by quantum state tomography, we measure the expectation values of the three spin projections $\sigma_x$, $\sigma_y$, $\sigma_z$. Only the projection onto $\ket0$ (i.e., $\sigma_z$) can be read out via fluorescence readout, so in order to obtain the other projections, the qubit has to be rotated accordingly before fluorescence readout using (rectangular) $\pi$ and $\frac{\pi}{2}$ pulses. The readout of each projection for each of the basis states is integrated over 10 million shots.
\item From the measured density matrices, we calculate the four matrices
\begin{align}
\rho_1^\prime &= \mathcal{E}\left(\ket{0}\bra{0}\right) \\
\rho_4^\prime &= \mathcal{E}\left(\ket{1}\bra{1}\right) \\
\rho_2^\prime &= \mathcal{E}\left(\ket{+}\bra{+}\right) \\ &- i\mathcal{E}\left(\ket{-}\bra{-}\right) - (1-i)(\rho_1^\prime + \rho_4^\prime)/2 \\
\rho_3^\prime &= \mathcal{E}\left(\ket{+}\bra{+}\right) \\ &+ i\mathcal{E}\left(\ket{-}\bra{-}\right) - (1+i)(\rho_1^\prime + \rho_4^\prime)/2
\end{align}
\item Finally, the process matrix $\chi$ is calculated from the $\rho_i^\prime$:
\begin{align}
\chi = \frac{1}{2} \begin{bmatrix}
I & \sigma_x \\
\sigma_x & -I
\end{bmatrix} \begin{bmatrix}
\rho_1^\prime & \rho_2^\prime \\
\rho_3^\prime & \rho_4^\prime
\end{bmatrix} \begin{bmatrix}
I & \sigma_x \\
\sigma_x & -I
\end{bmatrix}.
\end{align}
This is the matrix plotted for three cases in Fig.~\ref{fig3}.
\end{itemize}

\begin{widetext}
For the three cases shown in Fig.~\ref{fig3}, the numerical values of the measured $\chi$ matrices are as follows:
\begin{itemize}
\item[] \textbf{(a)} for correct ($\SI{100}{\percent}$) control amplitude and no detuning

\begin{align}\notag
\chi=\begin{pmatrix}
(0.001 + 0.000i) & (0.001 + 0.089i) & (-0.011 - 0.005i) & (0.002 -0.010i) \\
(0.001 -0.089i) & (0.991 + 0.000i) & (-0.002 -0.003i) & (-0.022 + 0.005i) \\
(-0.011 + 0.005i) & (-0.002 + 0.003i) & (-0.012 + 0.000i) & (0.001 -0.002i) \\
(0.002 + 0.010i) & (-0.022 -0.005i) & (0.001 + 0.002i) & (0.020 + 0.000i)
\end{pmatrix}
\end{align}

\item[] \textbf{(b)} for correct control amplitude and $\SI{8}{\MHz}$ detuning
\begin{align}\notag
\chi=\begin{pmatrix}
(0.005 + 0.000i) & (-0.050 + 0.009i) & (0.014 + 0.040i) & (-0.034 + 0.013i) \\
(-0.050  -0.009i) & (0.928 + 0.000i) & (0.034  -0.083i) & (0.237  -0.040i) \\
(0.014  -0.040i) & (0.034 + 0.083i) & (0.017 + 0.000i) & (-0.050 + 0.024i) \\
(-0.034  -0.013i) & (0.237 + 0.040i) & (-0.050  -0.024i) & (0.050 + 0.000i)
\end{pmatrix}
\end{align} 

\item[] \textbf{(c)} for $\SI{87.5}{\percent}$ of the correct control amplitude and no detuning

\begin{align}\notag
\chi=\begin{pmatrix}
(0.104 + 0.000i) & (-0.006 + 0.268i) & (0.017 + 0.012i) & (-0.000 + 0.004i) \\
(-0.006 + -0.268i) & (0.924 + 0.000i) & (0.000 + -0.015i) & (-0.020 + -0.012i) \\
(0.017 + -0.012i) & (0.000 + 0.015i) & (0.005 + 0.000i) & (-0.006 + 0.010i) \\
(-0.000 + -0.004i) & (-0.020 + 0.012i) & (-0.006 + -0.010i) & (-0.033 + 0.000i)
\end{pmatrix}
\end{align}

\end{itemize}
\end{widetext} 

\subsubsection*{Quantum process fidelity}
In order to quantify the fidelity of the experimentally realized processes in comparison to the ideally desired ones, we use the fidelity measure discussed in Ref.~\onlinecite{gilchrist_distance_2005} and used in Ref.~\onlinecite{howard_quantum_2006}: We start by expressing the experimental process matrix $\chi$ as a so-called process density matrix according to the Jamiolkowsk-formalism:
\begin{align}
\rho_{\mathcal E}=\left[\mathcal I \otimes \mathcal E \right]\left(\ket\Phi\bra\Phi\right),
\end{align}
where $\ket\Phi\bra\Phi$ is a projector onto a maximally entangled state in the Hilbert space of two qubits $\ket\Phi=\sum_j\ket j\ket j/\sqrt d$. The set of states $\{\ket j\}$ is an orthonormal basis set of the one-qubit space. For $d=2$, this can be rewritten as
\begin{align}
\rho_{\mathcal E}=\frac{1}{2}\sum_{ij}\ket i\bra j\otimes\mathcal E\left(\ket i\bra j\right).
\end{align}

For the ideal case of a perfect $\pi_x$ pulse we obtain
\begin{align}
\rho_{ideal} = \frac{1}{2}\begin{pmatrix}0&0&0&0\\0&1&1&0\\0&1&1&0\\0&0&0&0\end{pmatrix} = \ket{\psi}\bra{\psi},
\end{align}
i.e. the density matrix of the pure state $\ket{\psi}=(0,1,1,0)/\sqrt{2}$.

For our choice of basis operators ${A_i}$, the experimental process matrix $\chi$ is equal to the Choi matrix, which is proportional to the experimental process density matrix $\rho_{\mathcal E}$; hence we get
\begin{align}
\rho_{\mathcal E} = \frac{1}{d} \chi = \frac{1}{d} \sum_{i,j=0}^{d-1} \mathcal E \left(\ket{i}\bra{j}\right) \otimes \ket{i}\bra{j}.
\end{align}

Now we can evaluate the fidelity
\begin{align}
\mathcal F\left(\rho_{ideal}, \rho_{\mathcal E}\right)=\text{tr}\!\left(\sqrt{\sqrt{\rho_{ideal}}\rho_{\mathcal E}\sqrt{\rho_{ideal}}}\right)^2
\end{align}
which for a pure $\rho_{ideal}=\ket{\psi}\bra{\psi}$ simplifies to 
\begin{align}
\mathcal F\left(\ket{\psi}, \rho_{\mathcal E}\right)= \bra{\psi}\rho_{\mathcal E}\ket{\psi}.
\end{align}

This is the quantity quoted for three different realizations of a smooth optimal control $\pi_x$-pulse in Fig.~\ref{fig3}.

\subsubsection*{Physicality of estimated process}
Effects such as noise and finite sampling of the expectation values can lead to one or more eigenvalues of the process matrix $\chi$ to be negative, i.e. to the inference of an unphysical process from the data\cite{howard_quantum_2006}. To estimate the unphysicality of the inferred process, we first find a physically valid process matrix $\tilde\chi$ which is as close as possible to the measured process matrix $\chi$, but has real and positive eigenvalues only.

We define a $d^2 \times d^2$ complex, lower triangular matrix $T(t)$ with $d^4$ real parameters $t(i)$:
\begin{align}
T(t)=
\begin{pmatrix}
t_1&0&0&0\\
t_5+it_6&t_2&0&0\\
t_{11}+it_{12}&t_7+it_8&t_3&0\\
t_{15}+it_{16}&t_{13}+it_{14}&t_9+it_{10}&t_4
\end{pmatrix},
\end{align}

which can be used to form a completely positive matrix $\tilde\chi$:
\begin{align}
\tilde\chi=T^\dagger(t)T(t)
\end{align}

We numerically find the optimal $\tilde\chi$ by minimizing its deviation $\Delta$ from the measured $\chi$:
\begin{align}
\Delta\left(t(i)\right) = \sum_{mn}^{d^2}\left|\tilde\chi_{mn}(t)-\chi_{mn}\right|^2,
\end{align}
under the constraint of trace preservation
\begin{align}
\sum_{mn}^{d^2}\tilde\chi_{mn}(t)A_n^\dagger A_m=\mathbbm1.
\end{align}

Following Ref.~\onlinecite{howard_quantum_2006}, we calculate a number of measures of the difference between $\chi$ and $\tilde\chi$ to get an estimate of the unphysicality of $\chi$. Possible measures are the Frobenius norm $||X||_{Fro} = \sqrt{\mathrm{tr}\left[ X^{*} X \right]}$ of the difference matrix $X=\chi-\tilde\chi$ and the trace-distance $D\left( \chi, \tilde\chi\right)=\frac{1}{2} \mathrm{tr} \left[ \sqrt{(\chi-\tilde\chi)^\dagger (\chi-\tilde\chi)} \right] $, all of which are given in Table~\ref{tab:norm} for the three realizations of a smooth optimal control $\pi_x$ pulse for which process tomography was done: 

\begin{table}[tb]
\centering
\begin{tabular}[c]{l |c c c}
&$D$&$||X||_{Fro}$                                        \\
\hline
0 MHz detuning, 100\% control amplitude    &0.021    &0.025     \\
8 MHz detuning, 100\% control amplitude    &0.027    &0.032     \\
0 MHz detuning, 87.5\% control amplitude   &0.021    &0.024     \\
\end{tabular}
\caption{Estimates of the unphysicality of the measured process matrix $\chi$ for the three different processes shown in Fig.~\ref{fig3}.}
\label{tab:norm}
\end{table}

Note however, that these distance measures depend on the choice of the QPT basis operators $A_i$ and therefore are not comparable with results obtained using different basis operators. The values of the unphysicality norms for our data (Table~\ref{tab:norm}) are smaller than the ones obtained in Ref.~\onlinecite{howard_quantum_2006}. Their relatively small values (2 to \SI{3}{\percent} of the norm of the $\chi$ matrices) are an indication that our QPT experiments result in good reconstructions of the quantum process.

\section{State transfer fidelity: Comparison to composite pulses}
\begin{figure*}[tb]\label{fig_suppl_fidelities}
  \centering
    \includegraphics[width=1.0\textwidth]{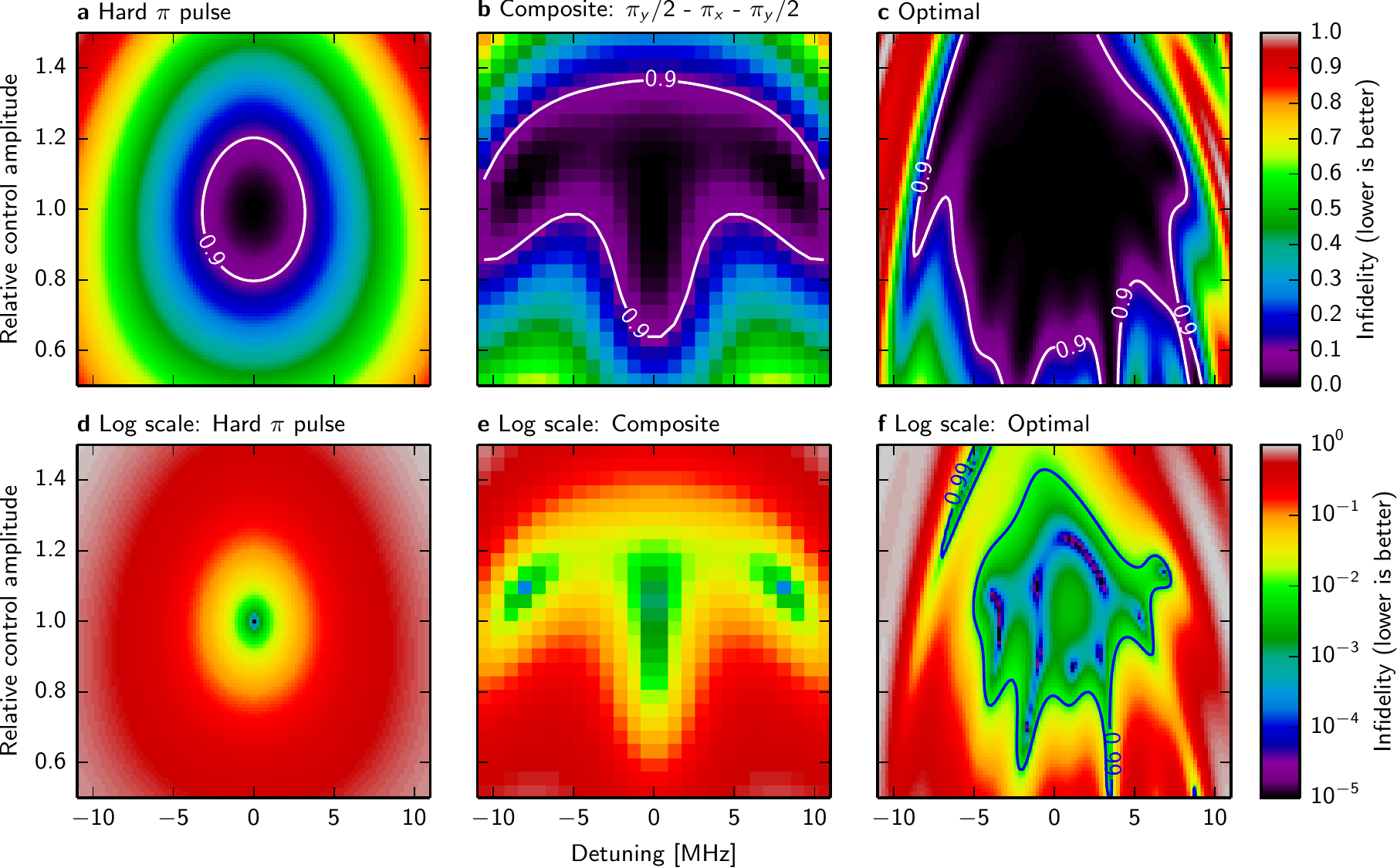}
  \caption{Comparison of simulated $\ket0$ to $\ket1$ transfer infidelity (i.e., one minus the expectation value of P(1) after the pulse) for \textbf{(a)} a hard $\pi_x$-pulse, \textbf{(b)} the composite hard pulse sequence $\left( \frac{\pi}{2} \right)_y - \pi_x - \left( \frac{\pi}{2} \right)_y$, and \textbf{(c)} a smooth optimal control pulse implementing the same operation. Upper row: linear, lower row: logarithmic colour scale. Lower values of the infidelity correspond to better performance. Contour lines as labelled in the plots. For (a) and (b) in the lower row, the \num{0.99} contour line collapses to less than one pixel and is therefore not shown.}
\end{figure*}

A great number of composite pulses (i.e., sequences of rectangular pulses) have been developed for nuclear magnetic resonance experiments, including many that are robust with respect to control amplitude deviations and detunings. While it is beyond the scope of this work to provide an in-depth comparison of these pulses to our smooth optimal control pulses, we examine one particular example: Ref.~\onlinecite{levitt_composite_2007} discusses the sequence $\left( \frac{\pi}{2} \right)_y - \pi_x - \left( \frac{\pi}{2} \right)_y$, which effectively results in a $\pi_x$ spin flip while offering good robustness properties (better results can be achieved with sequences of more than three pulses). In Fig.~\ref{fig_suppl_fidelities}, we plot the simulated $\ket0$ to $\ket1$ transfer infidelity (i.e., one minus the expectation value of $P(1)$ after the pulse) of this sequence in comparison to a hard $\pi_x$-pulse and a smooth optimal control pulse, on linear and logarithmic scales. The three control pulses are scaled to the same maximum control amplitude (Rabi frequency \SI{10}{\MHz}). Consequently, the simple hard pulse is shortest (\SI{50}{\ns}), followed by the composite sequence (\SI{100}{ns}), and the smooth optimal pulse (\SI{500}{\ns}).

Both on the linear and even more so on the logarithmic colour scale, the better performance of the smooth optimal control pulse is evident: The area of low infidelities encompasses much larger ranges of detunings and control amplitude inhomogeneities. Especially where very good (\SI{<1}{\percent} and \SI{<0.1}{\percent}) infidelities are required, smooth control pulses offer a clear advantage. Note that the smooth optimal control pulse shown here was designed for good fidelity over a large range of parameters for magnetometry purposes. Using the same control amplitude, modulation bandwidth and time budget, a considerably higher fidelity within a smaller window can be achieved.

\section{Evaluation of magnetometric sensitivity}
Following the protocols used and discussed in Refs.~\onlinecite{maze_nanoscale_2008,taylor_high-sensitivity_2008,dolde_electric-field_2011}, we measure the projection of an external AC magnetic field $B$ onto the NV symmetry axis by reading out the modulation of the spin echo amplitude caused by the field. We synchronise the free precession intervals of the spin echo sequence to the half-periods of the external field. Due to the Zeeman effect, the energies of the $|m_s| = 1$ spin sub-levels get shifted into opposite directions relative to the $m_s=0$ component during the first and second free precession periods, respectively. The resulting phase difference $\Delta \phi$ between the $m_s=0$ and $|m_s| = 1$ components of the precessing superposition state is converted to a difference in $\sigma_z$ population by the final $\frac{\pi}{2}$ pulse of the sequence. We read out the $\sigma_z$ population by exciting the center at \SI{532}{\nm} and recording the fluorescence intensity $I$, resulting in the magnetometric signal S:
\begin{align}
S = I = C N_{tot} \cos{\Delta \phi}.
\end{align}

For simplicity, we assume $B$ to be a square wave of amplitude $B_0$. The change in magnetometry signal $\mathrm{d}S$ due to a change of magnetic field amplitude $\delta B$ is the measured modulation $\Delta I$ in fluorescence photon counts:
\begin{align}
{\mathrm{d}S} &= \left| \frac{\partial S}{\partial B} \right| \delta B = \Delta I \\ &= \left| 2 N_{tot} C \tau \mu_B g \sin \left( 2 \tau \mu_B g B_0 \right) \right| \cdot \delta B \\ &\approx 2 N_{tot} C \tau \mu_B g \cdot \delta B.
\end{align}
Here, $C$ is the fluorescence readout contrast, $N_{tot}$ is the number of photons collected, $\tau$ the free precession time, $\mu_B$ the Bohr magneton and $g\approx2$ the NV center g-factor. Note that the readout contrast $C$ decreases as $\tau$ approaches the dephasing time $T_2$: Typically, $C \sim \exp \left[-(\tau/T_2)^n \right]$, with $n \approx 0.5 \ldots 2$ (cf. Ref.~\onlinecite{stanwix_coherence_2010}). Since we are interested in finding the largest change in signal, and hence the best sensitivity, we evaluate the change in signal at its steepest slope, and hence we set the sine equal to one in the last simplification.

We define the smallest detectable magnetic field $\delta B_{min}$ as the signal that is just as large as the noise, i.e. at a signal to noise ratio $\mathrm{SNR}=1$. The noise is  limited by the photon shot noise, $\sqrt{N_{tot}}$. The total number of photons collected in a magnetometry measurement depends on the peak counts per second ($N_{cps}$) generated by the detection optics and electronics, the total integration time $T$, and the ratio of per-shot signal acquisition time $t_{acq}$ ($\SI{200}{ns}$ in our case) to the total duration of one shot (which consists of preparation and free precession phases):
\begin{align}
N_{tot} = N_{cps} \frac{t_{acq}}{2 \tau + t_{preparation}} T
\end{align}

Hence, the minimum detectable magnetic field is inversely proportional to the square root of the total photon number, which rises linearly with the total integration time $T$:
\begin{align}
\delta B_{min} = \frac{1}{N_{tot} C \tau \mu_B g \sqrt{N_{tot}}}
\end{align}

Multiplying out the root of the integration time yields the sensitivity in units of $\mathrm{T/\sqrt{Hz}}$. 

To measure the sensitivity, we scan the amplitude of an applied magnetic field and record the sinusoidal modulation of the fluorescence intensity. We fit the oscillation, determine the steepest slope, and compare it to the noise on the oscillation (taken as the standard deviation of the fit residuals). The sensitivity is the change in signal that is just as large as the noise.

The photon count rate and the readout contrast -- and hence the sensitivity -- depend on the excitation laser power and the choice of colour filters placed in the fluorescence path, as well as the chemical structure of the host material, which has to be carefully optimized. The excitation power clearly affects the fluorescence rate, but also the dynamic equilibrium between the negative (\nvmnosp) and neutral (\nvznosp) charge states of the center. Only \nvm fluorescence contributes to the magnetometric signal, while \nvz is insensitive to magnetic fields and contributes to the background. The fluorescence spectra of the two charge states largely overlap, so careful choice of the fluorescence filter passband is required to find an optimal trade-off between signal strength and contrast. We found that a \SI{750}{\nm} long pass filter maximizes the sensitivity. In our experimental realization, it was necessary to use rather low excitation power (\SI{50}{\micro \watt}) due to the limited dynamic range of our detectors. This lowers the number of fluorescence photons emitted and hence the sensitivity. Since this limitation would be easy to overcome by using more suitable detectors, we correct for its effect during data analysis.

As illustrated in Fig.~\ref{fig4}, the sensitivity drops significantly when the control pulses are detuned from resonance for a conventional, hard pulse (for example, due to a spatial or temporal inhomogeneity of a bias field). This is due to the fact that a detuning leads to control errors and imperfect $\pi$ and $\frac{\pi}{2}$ rotations in the spin echo sequence. Consequently, the position of steepest slope of the signal gets shifted. In a practical measurement scenario, this effect however would be unknown and contribute to a loss in sensitivity. To realistically quantify the sensitivity, we therefore have to evaluate the slope at the position of steepest slope for the undetuned case. Similarly, a temporal or spatial change or inhomogeneity in control amplitude (over the integration time of the measurement, or measurement area) worsens sensitivity, unless alleviated by robust control methods. We note that the symmetric nature of the spin echo sequence already results in some degree of robustness with respect to these control errors, even when using rectangular pulses (likely more so than the Ramsey sequences typically used in DC magnetometry).

\bibliography{smooth_control_paper_arxiv}
\end{document}